\documentclass{pasj00}
\begin{document}
%\SetRunningHead{Author(s) in page-head}{Running Head}
\SetRunningHead{Sugizaki et al.}{In-Orbit Performance of MAXI/GSC}

%\Received{}%{yyyy/mm/dd}
%\Accepted{}%{yyyy/mm/dd}
%\Published{}%{yyyy/mm/dd}

\title{In-Orbit Performance of MAXI Gas Slit Camera (GSC) on ISS}

%%% begin:list of authors
% Do NOT capitalize all letters in "textsc".
\author{
Mutsumi \textsc{Sugizaki} \altaffilmark{1}, 
Tatehiro \textsc{Mihara} \altaffilmark{1}, 
Motoko \textsc{Serino} \altaffilmark{1},
Takayuki \textsc{Yamamoto} \altaffilmark{1},
Masaru \textsc{Matsuoka} \altaffilmark{1,2}, 
Mitsuhiro \textsc{Kohama} \altaffilmark{2}, 
Hiroshi \textsc{Tomida} \altaffilmark{2}, 
Shiro \textsc{Ueno} \altaffilmark{2}, 
Nobuyuki \textsc{Kawai} \altaffilmark{3}, 
Mikio \textsc{Morii} \altaffilmark{3},
Kousuke \textsc{Sugimori} \altaffilmark{3},
Satoshi \textsc{Nakahira} \altaffilmark{4},
Kazutaka \textsc{Yamaoka} \altaffilmark{4},
Atsumasa \textsc{Yoshida} \altaffilmark{4},
Motoki \textsc{Nakajima} \altaffilmark{5},
Hitoshi \textsc{Negoro}\altaffilmark{5}, 
Satoshi \textsc{Eguchi} \altaffilmark{6},
Naoki \textsc{Isobe} \altaffilmark{6},
Yoshihiro \textsc{Ueda} \altaffilmark{6},
and 
Hiroshi \textsc{Tsunemi} \altaffilmark{7}
}

\altaffiltext{1}{MAXI team, RIKEN, 2-1 Hirosawa, Wako, Saitama 351-0198} 
\email{sugizaki@riken.jp} 
\altaffiltext{2}{ISS Science
  Project Office, ISAS, JAXA, 2-1-1 Sengen, Tsukuba, Ibaraki 305-8505}
\altaffiltext{3}{Department of Physics, Tokyo Institute of Technology,
  2-12-1 Ookayama, Meguro-ku, Tokyo 152-8551}
\altaffiltext{4}{Department of Physics and Mathematics, Aoyama Gakuin
  University, 5-10-1 Fuchinobe, Sagamihara, Kanagawa 229-8558}
\altaffiltext{5}{Department of Physics, Nihon University, 1-8-14,
  Kanda-Surugadai, Chiyoda-ku, Tokyo 101-8308}
\altaffiltext{6}{Department of Astronomy, Kyoto University,
  Oiwake-cho, Sakyo-ku, Kyoto 606-8502, Japan}
\altaffiltext{7}{Department of Earth and Space Science, Osaka
  University, 1-1 Machikaneyama, Toyonaka, Osaka 560-0043, Japan}

%\email{ccccc@xxx.xxx.xx.xx}

%\author{B-Firstname \textsc{B-Familyname}}
%\affil{B-Address of Institute}\email{bbbbb@xxx.xxx.xx.xx}
%\and
%\author{C-Firstname {\sc C-Familyname}}
%\affil{C-Address of Institute}\email{ccccc@xxx.xxx.xx.xx}

%%% end:list of authors

%%% Please use the following style in case that sorting by 
%%% affilation is impossible. 
%
% \author{%
%   D-Firstname \textsc{D-Familyname}\altaffilmark{1}
%   E-Firstname \textsc{E-Familyname}\altaffilmark{1,2}
%   and
%   F-Firstname \textsc{F-Familyname}\altaffilmark{2}}
% \altaffiltext{1}{Address of Institute}
% \email{ddddd@xxx.xxx.xx.xx}
% \email{eeeee@xxx.xxx.xx.xx}
% \altaffiltext{2}{Address of Institute}

%% `\KeyWords{}' always has to be placed before `\maketitle'.
 %Do NOT move this preamble from here!

\maketitle

\begin{abstract}
  We report the in-orbit performance of the Gas Slit Camera (GSC) on
  the MAXI (Monitor of All-sky X-ray Image) mission carried on the
  International Space Station (ISS).  Its commissioning operation
  started on August 8, 2009, confirmed the basic performances of the
  effective area in the energy band of 2--30 keV, the spatial
  resolution of the slit-and-slat collimator and detector with
  1.5$^\circ$ FWHM, the source visibility of 40-150 seconds for each
  scan cycle, and the sky coverage of 85\% per 92-minute orbital
  period and 95\% per day.  The gas gains and read-out amplifier gains
  have been stable within 1\%.  The background rate is
  consistent with the past X-ray experiments operated at the similar
  low-earth orbit if its relation with the geomagnetic cutoff rigidity
  is extrapolated to the high latitude.  We also present the status of
  the in-orbit operation and the calibration of the effective area and
  the energy response matrix using Crab-nebula data.

\end{abstract}
%\linenumbers

\section{Introduction}

MAXI (Monitor of All-sky X-ray Image) is the first astronomical
mission operated on the ISS (International Space Station)
(\cite{matsuoka_pasj2009}).  The payload was delivered to the ISS by
Space Shuttle Endeavour on July 16, 2009, and installed on the
Japanese Experiment Module -- Exposed Facility (JEM-EF or Kibo-EF) on
July 24.  The mission was designed to achieve the best sensitivity and
the highest energy resolution among the all-sky X-ray monitors
performed so far by using two kinds of X-ray slit cameras, Gas Slit
Camera (GSC, \cite{mihara_paper1}) and Solid-state Slit Camera (SSC,
\cite{tomida_ssc1}), which work complementary.

\begin{figure*}
  \begin{center}
    %\FigureFile(5.818cm,){col_scheme5.eps}
    \FigureFile(5.818cm,){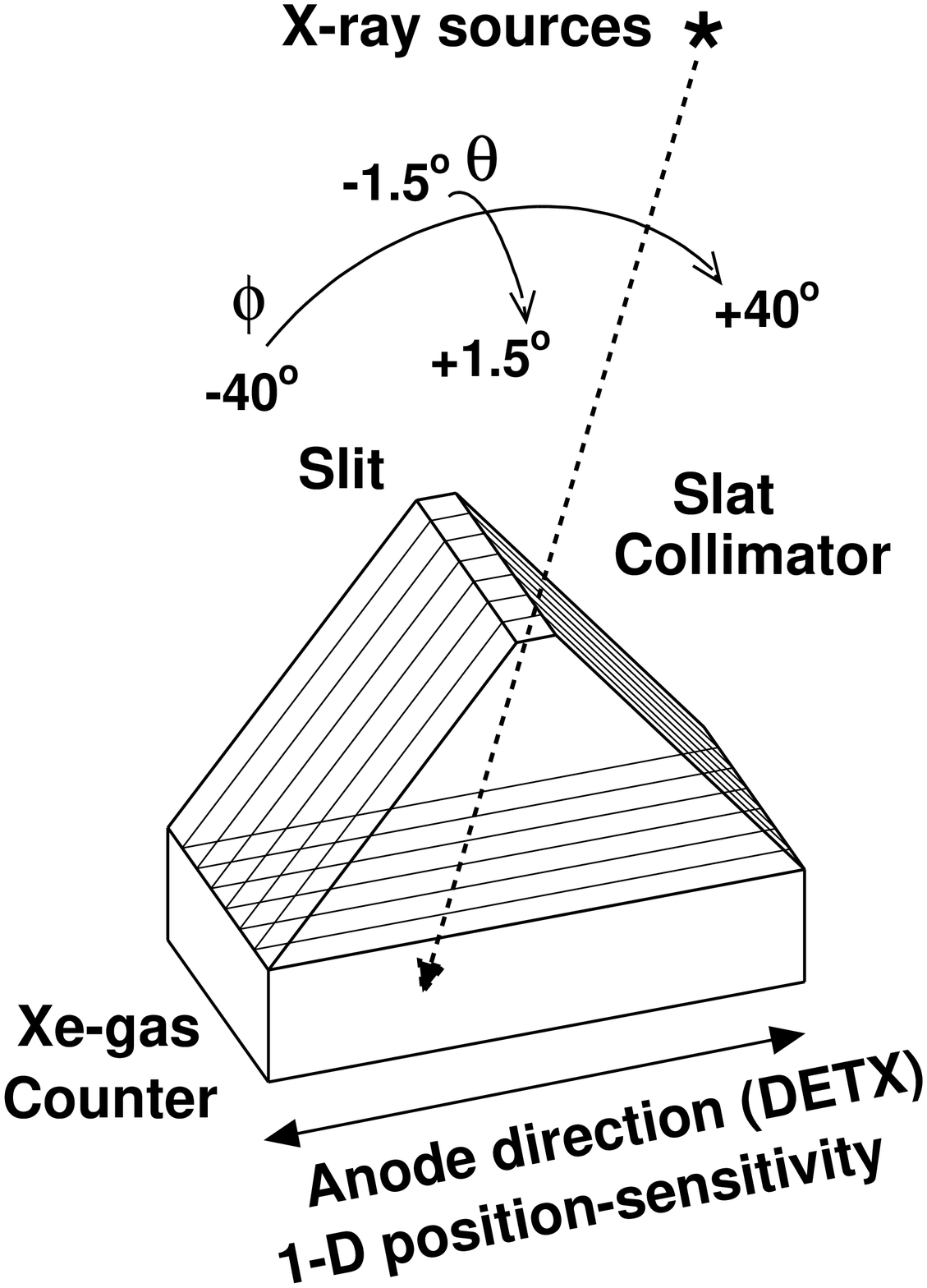}
    \hspace{5mm}
    %\FigureFile(8.338cm,){maxi_iss_move_bw5.eps}
    \FigureFile(8.338cm,){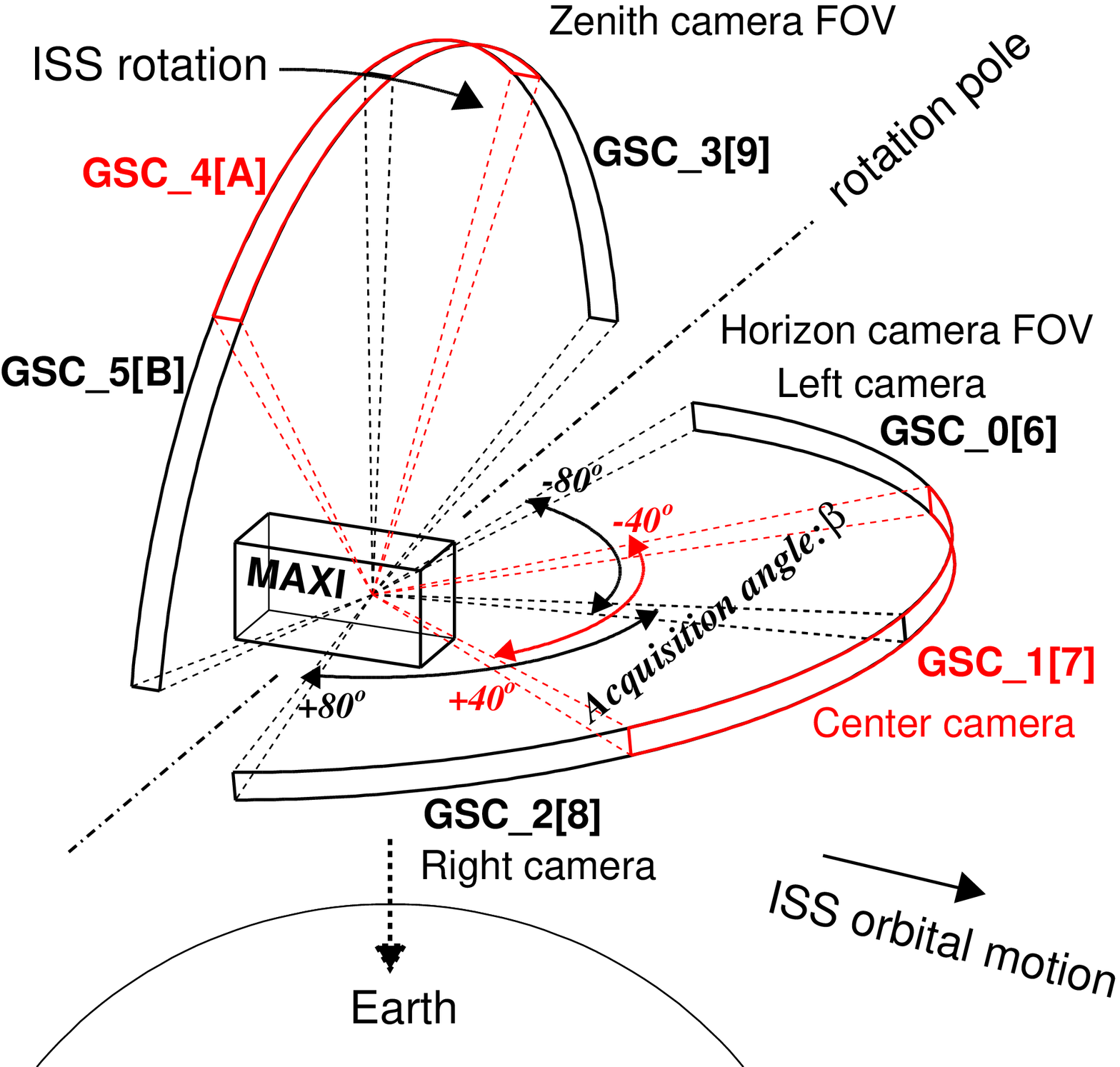}
  \end{center}
    \caption{ 
      (Left): Schematic view of a single GSC camera unit and
      its field of view (FOV) of $1.5^\circ {\rm (FWHM)} \times
      80^\circ$. (Right): FOVs of the six camera units including
      12 counters (GSC\_0, ..., GSC\_9, GSC\_A, GSC\_B) on
      the ISS low-earth orbit.  Every two counters, 
      such as GSC\_0 and GSC\_6, are used as a pair for a
      redundancy.  Left, center, and right camera units 
      cover the area of source-acquisition angle $\beta$ of
      $-80^\circ<\beta<0^\circ$, $-40^\circ<\beta<+40^\circ$, and
      $0^\circ<\beta<+80^\circ$, respectively.  
    }
    \label{fig:maxi_gsc_schematic}
  
\end{figure*}

The GSC is the main X-ray camera to cover the energy band from 2 to 30
keV with a large area.  It employs conventional slit camera that
consists of Xe-gas proportional counters and slit-and-slat
collimators.  Twelve gas counters achieve a large detector area of
5350 cm$^2$. They are assembled into six identical units to cover wide
fields of views (FOVs) of $1.5\times160$ degree$^2$.  They are
embodied in the MAXI payload module such that they cover the
earth-horizon and the zenith directions with an equal area.  It
enables us to cover the entire sky every 92-minute orbital period even
if the GSC has to be partly downed in a heavy particle-radiation
environment such as the passage of the SAA (South Atlantic Anomaly).
The Xe-gas counters employ resistive carbon-wire anodes to acquire
one-dimensional position sensitivity.  The gas volume of each counter
is separated into six carbon-anode cells for main X-ray detectors and
ten tungsten-anode cells for veto counters. The veto cells surround
the carbon-anode cells to reduce backgrounds by the
anti-coincidence-hit logic. Total 14 anode signals in each counter are
read out independently through an electronics unit and processed by
on-board data processors.  
Figure \ref{fig:maxi_gsc_schematic}
illustrates the schematic views of a single GSC camera unit and FOVs of the
whole six units on the ISS low-earth orbit.  
The details of the
instruments are described in \citet{mihara_paper1}.

This paper presents the GSC in-orbit performance including the
operation and calibration status. The SSC in-orbit performance
is presented in \citet{tsunemi_ssc2}.

\section{Operation}

\subsection{Initial Operation}

The main power of the MAXI payload is supplied from the JEM-EF via the
Payload Interface Unit (PIU).  It was activated on August 3, one week
later since the payload was installed on the JEM-EF.  On August 4, the
GSC electronics units (GSC-E) including the Radiation-Belt Monitor
(RBM) were switched on.  The on-board RBM parameters, thresholds of
the lower discriminators and count-rate criteria to down the gas
counters, were then set up using the data taken for the first three
days.  On August 8, the high voltage of the first gas counter 
was activated and increased up to the nominal operation voltage of
1650 V.  The first test observation was then performed for three hours
and various analog/digital-electronic functions were verified.  All
the 12 counters were activated at the nominal 1650 V since August 10
to 13 one by one.

Each gas counter has 14 anode signals for both two ends of six
carbon-wire anodes and two veto signals.  Each anode signal has
individual pre-amplifier, a shaping amplifier, a lower discriminator
(LD), and an Analog-to-Digital Converter (ADC) which produces 14-bit
pulse height.  The gain of the shaping amplifier and the threshold of
the lower discriminator of each signal can be changed by command.
These parameters are adjusted according to the gas gain of individual
anode wire.  The first all-sky image by all 12 counters was taken on
August 15.

\subsection{RBM}

GSC incorporates Radiation-Belt Monitor (RBM) to monitor the flux of
the cosmic rays in orbit.  Two silicon-PIN-diode detectors with 0.2-mm
thickness and 5$\times$5 mm$^2$ area are equipped with each of GSC
horizon and zenith center units.  The detector on the horizon unit
(RBM-H) is faced to the tangential direction of the ISS motion along
the earth horizon. The zenith unit (RBM-Z) is faced perpendicular to
the horizontal plane.  The RBMs always monitor count rates of cosmic
rays with a certain energy deposit, whose threshold can be changed by
commands.  The level of the threshold is set at 50 keV, which
corresponds to the half of the minimum ionization-loss energy of the
relativistic particles.

\begin{figure}
  \begin{center}
    %\FigureFile(8.5cm,){gea_rbmmap_sum_hsv4.eps}
    %\FigureFile(8.5cm,){geb_rbmmap_sum_hsv4.eps}
    %\FigureFile(8.5cm,){gbm_rbmmap_ratio.eps}
    \FigureFile(8.5cm,){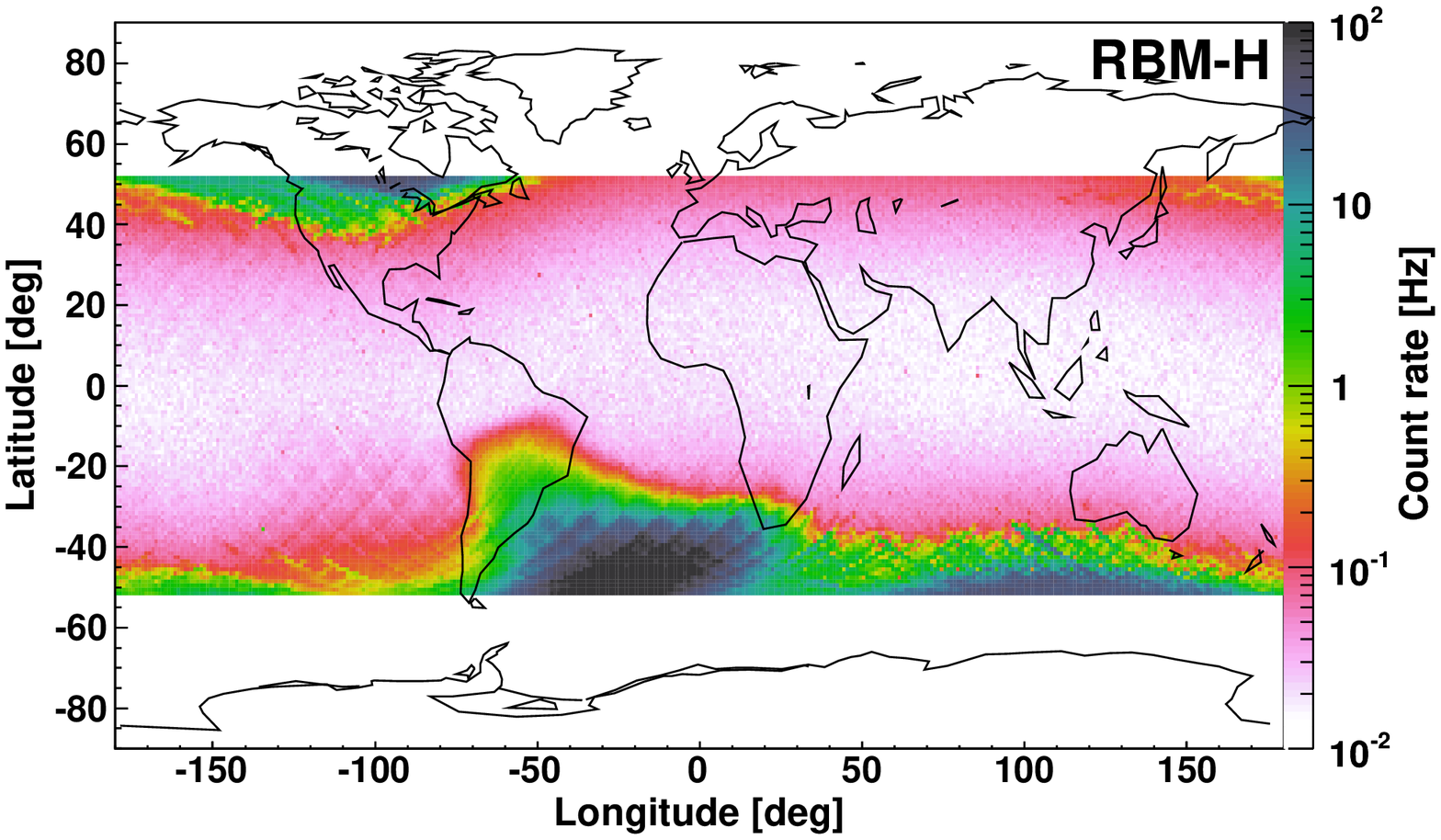}
    \FigureFile(8.5cm,){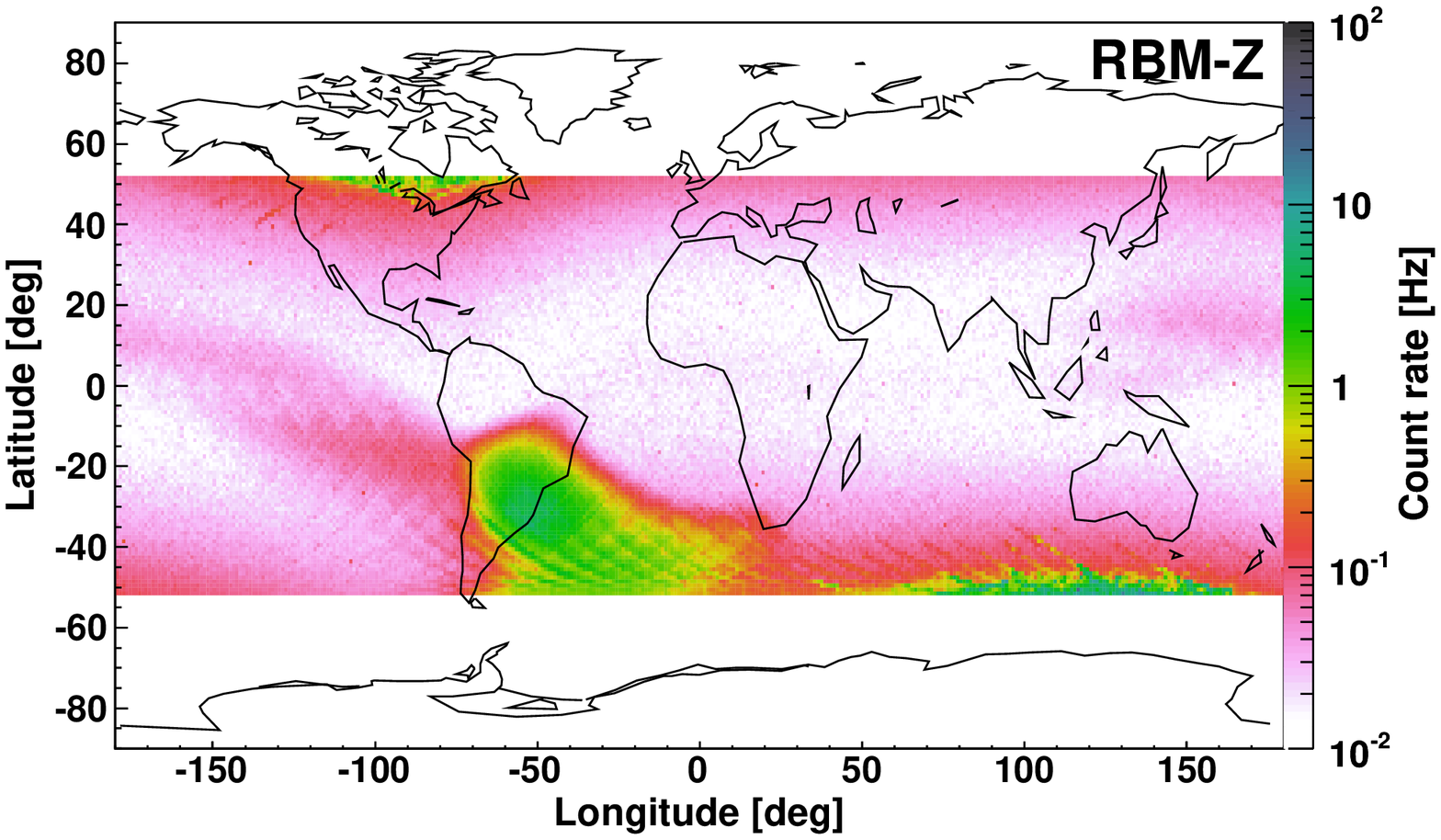}
    \FigureFile(8.5cm,){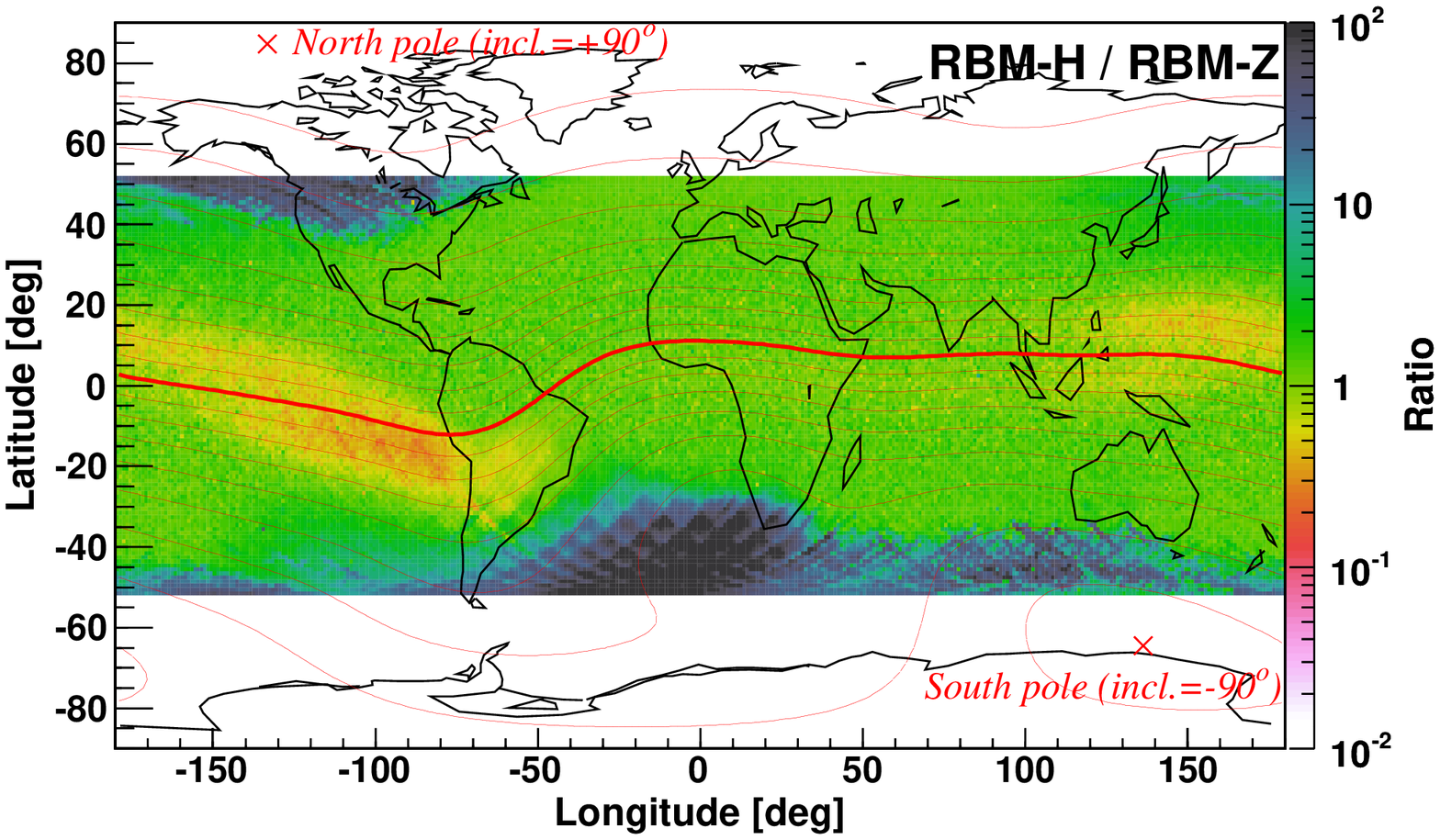}
  \end{center}
  \caption{RBM count-rate maps of horizon (RBM-H, top panel) and
    zenith (RBM-Z, middle panel) units.  Bottom panel shows their
    ratio (RBM-H/RBM-Z) and contours of the inclination angle of
    the magnetic fields from the horizontal plane
    are drawn by $10^\circ$ step from $-90^\circ$ (south pole) 
    to $90^\circ$ (north pole). 
  }
  \label{fig:rbmmap}
\end{figure}

Figure \ref{fig:rbmmap} shows the RBM count-rate maps on the ISS
orbital area by RBM-H and RBM-Z units, respectively, averaged over the
data taken for the first year.  
The ratio of the RBM-H to the RBM-Z
count rate is shown at the bottom where contours of inclination angle
of the magnetic field from the earth-horizontal plane are overlaid.
The two RBM rates agree with each other in the low count-rate area
below 0.2 Hz, where the anti-correlations with the geomagnetic cutoff
rigidity (COR) are clearly seen.  It indicates that the origins are
energetic cosmic rays.  The cosmic-ray flux $\sim$ 1 counts cm$^{-2}$
s$^{-1}$ and its relation with the COR are consistent with those
obtained by ASCA (\cite{makishima_pasj1996}) and Suzaku
(\cite{kokubun_pasj2007}) in low-earth orbit although the altitude of
the ISS, 340--360 km, is different from that of these satellite,
530--590 km.

While the two RBM rates are agreed at the low radiation area, they are
quite different in the high radiation area around the SAA and the both
geomagnetic poles located at the north of the north American continent
and the south of the Australian continent. The RBM-H rate is higher by
an order of magnitude than the RBM-Z rate in these area.  From the
correlation between their ratio and the inclination of the magnetic
field, the difference is considered to be due to the anisotropy of the
flux of trapped particles.  A large number of cosmic-ray particles
trapped by the geomagnetic fields are moving circularly on the
horizontal plane around the magnetic pole.  The RBM-Z detector faced
on the zenith direction is insensitive to the particles moving on the
horizontal plane.  The anisotropy of the trapped particles at the high
latitude has also been recognized by images of particle tracks taken
by the SSC CCD imager (\cite{tsunemi_ssc2}).

\subsection{High Voltage Operation}
\label{sec:HVoperation}

We started full GSC operation with twelve counter units, 
GSC\_0, ..., GSC\_9, GSC\_A, GSC\_B, on August 15, 2009.  The high
voltages of all the twelve gas counters were set at the nominal 1650
V.  They are reduced to 0 V when the ISS passes through heavy particle
background area defined by an on-board Radiation-Zone (RZ) map.  In
the beginning of the mission, we set the RZ map only at the SAA.

On September 8, 2009, the analog power on the GSC\_6 counter was
suddenly down. It was followed by the power down of another counter,
GSC\_9, on September 14.  The data indicate that a carbon-anode wire
in each of these counters was fractured.  It was also suggested that
the location fractured on the carbon wire would be where a large
amount of discharges occurred repeatedly.

We then changed the counter operation strategy so that any risks to
cause potential damages on the carbon-anode wire such as discharge and
heavy irradiation should be avoided as much as possible. The
high-voltage reduction at the high latitude above 40$^\circ$ was
employed since September 23.  Two noisy counters, GSC\_A and GSC\_B,
which are considered to be fractured sooner, were also tentatively
stopped on September 26.

On March 26, 2010, the analog power on the GSC\_3 counter was downed,
which were considered to be due to a carbon-wire fracture by breakdown
discharge again.  Since then, the operation voltage was reduced to
1550 V if the counter had any discharge experience.

As the results of these limited HV operation, the effective
observation efficiency is reduced to about 40\%.  It is expected to
reduce the sensitivity.  The three counters suffering from the
breakdown will be soon activated with great care.

\section{In-Orbit Performance}

\subsection{Background}

The GSC counter has a one-dimensional position sensitivity and
performs scanning observation.  Backgrounds for any X-ray sources on
the sky can be roughly estimated from the levels of the adjacent
source-free region.  The studies of the backgrounds are important
because the residual backgrounds and its uncertainties limit the
source-detection sensitivity.

\subsubsection{LD-hit Rate and Processed Event Rate}

Each GSC gas counter embodies six carbon-anode cells for main X-ray
detector and the veto cells surrounding the carbon-anode cells.  The
cosmic-ray backgrounds are screened out on board by applying an
anti-coincidence matrix to hit patterns of all anode signals in each
counter.  The efficiency of the background reduction depends on the LD
threshold of the read-out signals.  The amplifier gains and LD levels
of all the read-out signals are tuned during the initial verification
operation.

\begin{figure}
  \begin{center}
    %\FigureFile(8.5cm,){evrigi_gsc0_ut20090818.eps}
    \FigureFile(8.5cm,){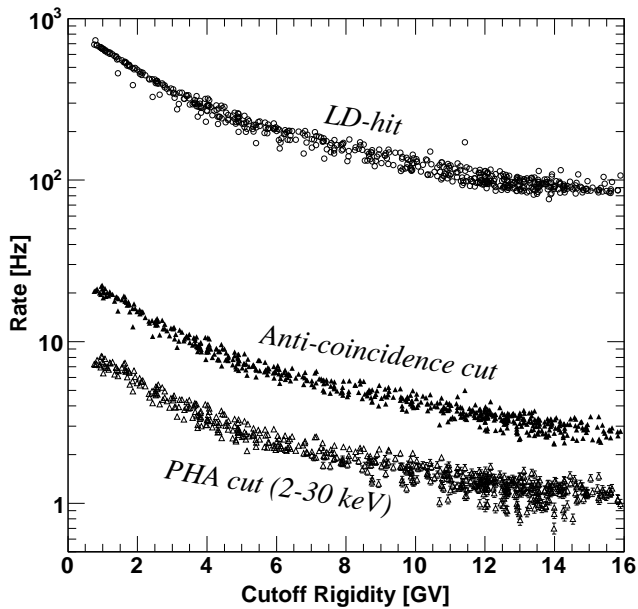}
  \end{center}
    \caption{
      GSC LD-hit rate, on-board processed event rate after
      anti-coincidence cut, and PHA-filtered event rate with the
      energy band of 2--30 keV per anode cell against geomagnetic 
      cutoff rigidity.
    }
    \label{fig:evbgrate_rigidity}
\end{figure}

The background rates of all the 12 GSC counters are found to have a
similar time variation during the entire observation-operation period.
They exhibit clear anti-correlations with the geomagnetic COR in
orbit.  Figure \ref{fig:evbgrate_rigidity} illustrates the LD-hit
rate, the processed event rate after the anti-coincidence cut, and the
cleaned event rate with the PHA (Pulse-Height Amplitude) range of the
2--30 keV energy band against the COR.  Each data point represents the
rate per anode averaged over six anodes in a counter, GSC\_0, every
minute, taken for a day on August 18, 2009, except while the ISS is in
a heavy particle-irradiation area detected by RBM and RZ map.  Data
expected to include significant X-ray events from celestial bright
X-ray sources are also excluded.  The ratio of the
anti-coincidence-event rate to the LD-hit rate is about 1:30 and
almost constant over the entire period. It is quite similar with the
Ginga-LAC case (\cite{turner_ginga_lac}).  The LD-hit rate is well
agreed with the RBM rate taking the detector area, 90 cm$^2$ per
anode, into account. It indicates that they are the energetic
cosmic-ray origin.

\subsubsection{Energy Spectrum}

Figure \ref{fig:bgd_spec_raw} shows energy spectra of the residual
backgrounds sorted by COR and the estimated contribution of the
unresolved Cosmic X-ray Background (CXB) within the FOV
(e.g. \cite{1995PASJ...47L...5G}).  The profile shows the fluorescent
lines of Ti-K at 4.6 keV from the gas-counter body and Cu-K at 8.1 keV
from the slat collimator made of phosphor bronze.  These emission
lines are used for the energy response calibration.  The shape of the
continuum in the energy band above 8 keV is quite sensitive to the
COR. It suggests that the origins of these components would be
energetic cosmic rays.  The CXB component is expected to be dominant
in the energy band below 5 keV at COR$>$8.

Figure \ref{fig:bgd_spec_uf} illustrates the background spectrum
normalized by effective area, and those of the Ginga LAC
(\cite{hayashida_ginga_bgd}) and the RXTE PCA (\cite{jahoda_xte_pca}),
where expected contribution of the CXB in each instrument are
included.  Crab-like source spectra are shown together as the
comparison.  The GSC background is approximately 2 mCrab at 4 keV and
10 mCrab at 10 keV.  The level is almost comparable to that of the
Ginga LAC and slightly higher than that of the RXTE PCA.

\begin{figure}
  \begin{center}
    %\FigureFile(8.5cm,){bgdspec_20090818_v2.eps} 
    \FigureFile(8.5cm,){fig4.eps} 
  \end{center}
    \caption{ GSC background spectra per anode cell sorted by COR 
      (Cutoff Rigidity) and
      estimated contribution of the CXB (Cosmic X-ray Background).
      The data are average spectra of the entire 12 counters taken for
      a day on August 18, 2009.  Ti-K and Cu-K fluorescent lines are
      seen at 4.6 keV and 8.1 keV, respectively.}
    \label{fig:bgd_spec_raw}

  \begin{center}
    %\FigureFile(8.5cm,){bgdspec_sens3_ph.eps} 
    \FigureFile(8.5cm,){fig5.eps} 
  \end{center}
    \caption{GSC background energy spectrum normalized  by effective area
      and comparison with those of Ginga-LAC, RXTE-PCA, and
      Crab-like source spectra.  Expected contributions of the
      CXB in each instrument are included.  }
    \label{fig:bgd_spec_uf}
\end{figure}

\subsubsection{Spatial Distribution}

\begin{figure}
  \begin{center}
    %\FigureFile(8.5cm,){detxprof0345_pi40_200_20090817.eps}
    \FigureFile(8.5cm,){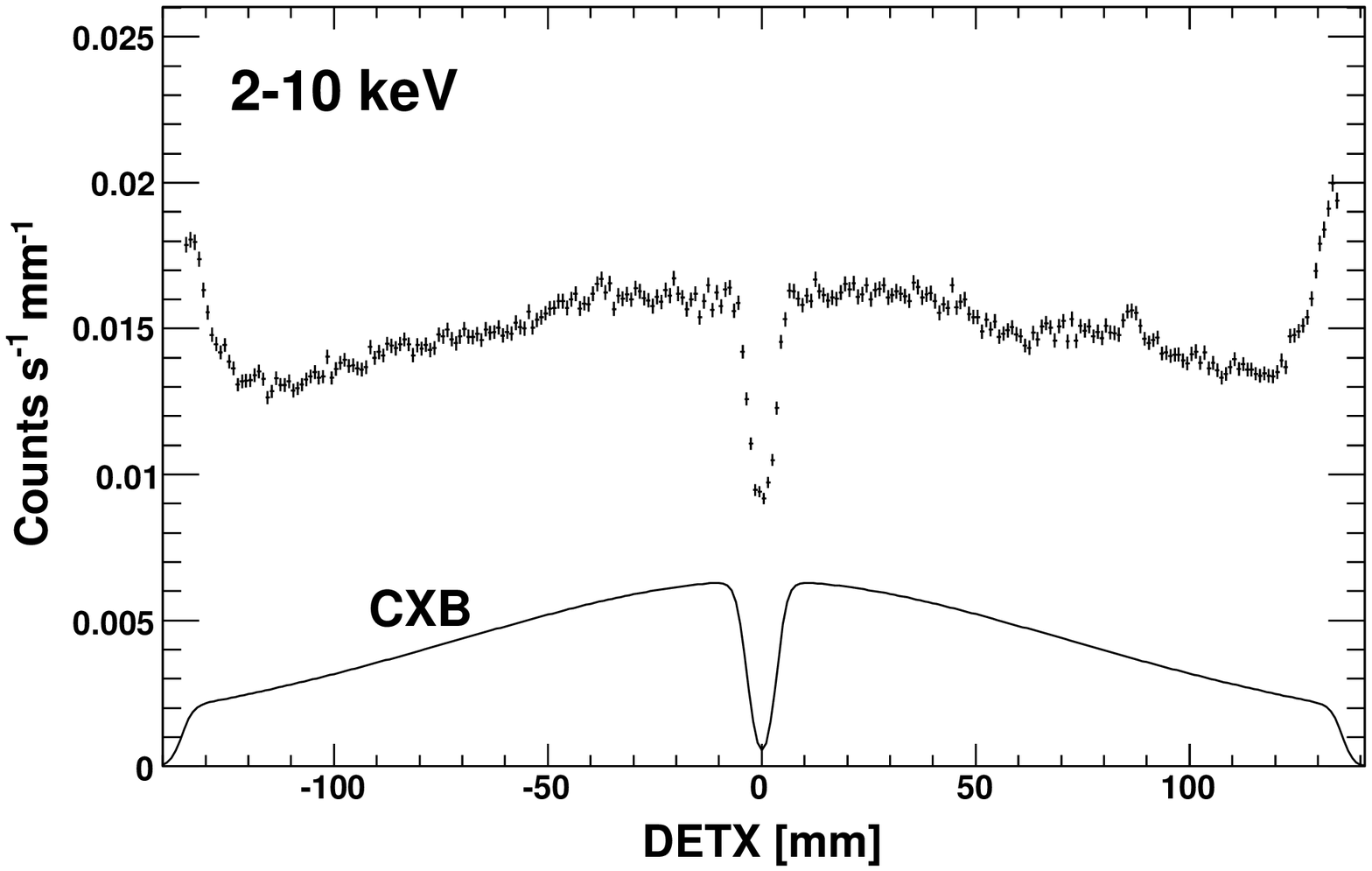}

    \vspace{2mm}

    %\FigureFile(8.5cm,){detxprof0345_pi200_600_20090817.eps}
    \FigureFile(8.5cm,){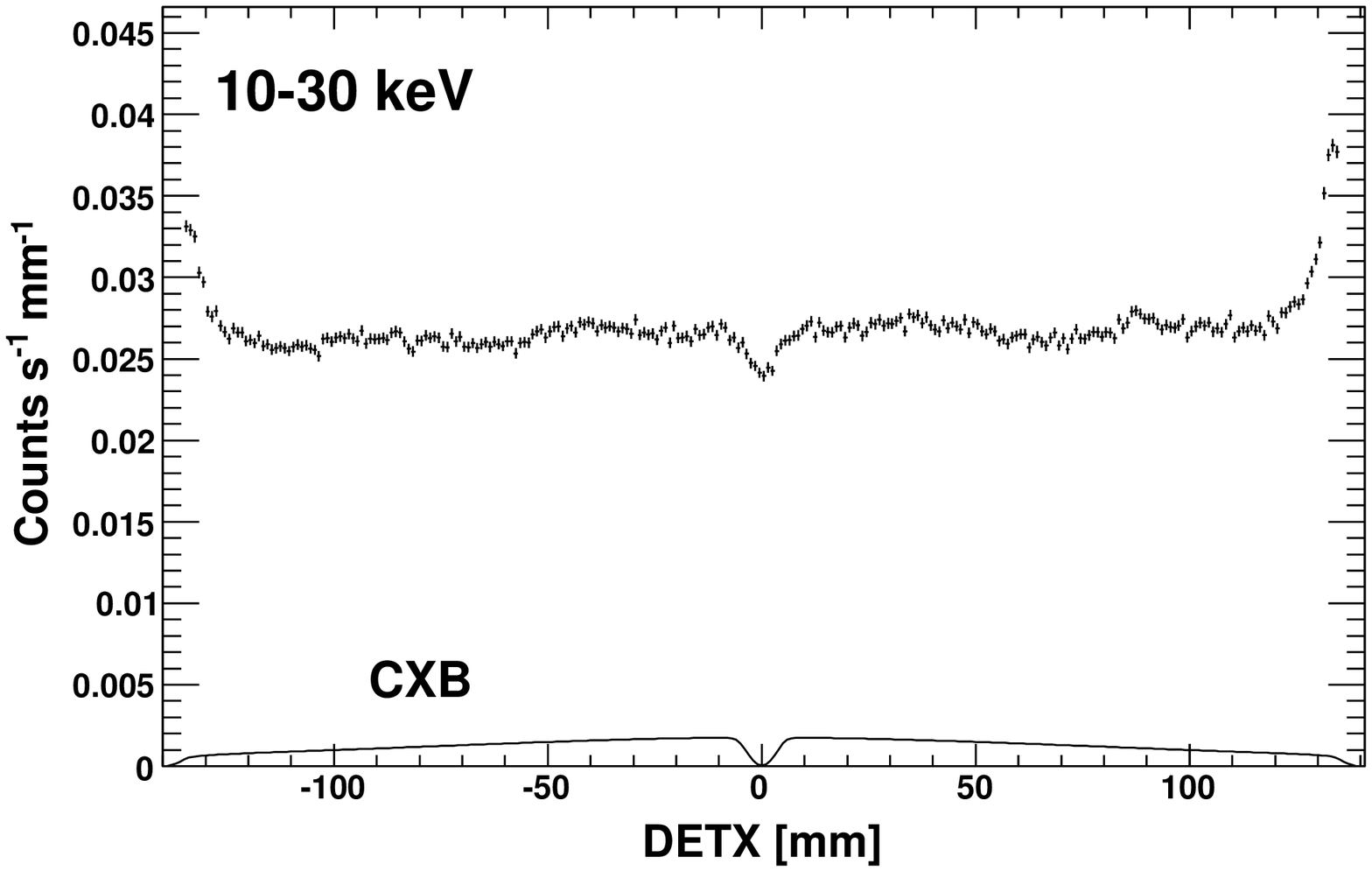}

  \end{center}
    \caption{ Background spatial distribution in GSC gas counters
      along anode wires (DETX) in each energy band of 2-10 keV (top)
      and 10-30 keV (bottom).  Estimated contributions of the CXBs are
      shown with solid lines.  The dips at the center (DETX$=$0) are
      due to the shadow of the support structure of the counter
      beryllium window.  }
    \label{fig:bgd_detx}
\end{figure}

Figure \ref{fig:bgd_detx} shows spatial distributions of the residual
backgrounds along the anode wire in 2-10 and 10-30 keV energy bands
and the estimated contributions of the CXB.  Each GSC counter has a
sensitivity gap at the center of the anode wire due to the frame
structure to support 0.1-mm thick beryllium window. It causes the dip
at the center in the spatial distribution.  The depth of the dips can
be mostly explained by the extinction of the CXB component at the
support structure.

The profile of the residual background left after subtracting the CXB
component is primarily flat and slightly increases at both ends of the
anode wires.  It is because the efficiency of the anti-coincidence
reduction decreases at the counter edge.  The sensitivity is expected
to get worse in these area.

\subsection{Energy Response and Gain Stability}

Each gas counter is equipped with a calibration source embodying
radioactive isotope $^{55}$Fe at the end of one of six carbon anodes
(the left-side end of carbon anode \#2). It always exposes collimated
X-ray beam with a diameter of 0.2 mm.  The counter gas gain and energy
resolution have been continuously monitored with the 5.9-keV line
since the ground calibration test.  The results show that the gas gain
is stable within 2\%, which agrees well with the results of the
vacuum-environment test on the ground.  The variation in orbit during
the first year was within 0.8\%.  The energy resolution of 18\% at 5.9
keV was stable within the statistical uncertainty.

The energy scale of the PHA of all the read-out signals are monitored
at the entire location along the carbon-anode wire using Ti-K and Cu-K
fluorescent lines in the background, as seen in figure
\ref{fig:bgd_spec_raw}.  The time variations for the first year was
within 1\% in the entire counter area.  This information is
employed for a PHA-to-PI (Pulse-height Invariant) conversion in the
data reduction process.

\subsection{Detector Position Response and Stability}

The incident X-ray position along a carbon-anode wire is calculated
from the PHA ratio of two read-out signals to the both anode ends.
All the required calibration data to derive the detector position from
the PHA ratio were collected in the ground tests
(\cite{mihara_paper1}).

The stability of the detector position response has been monitored
with the calibration source since the ground calibration tests.  The
time variation of the calculated position during the first year was
less than 0.2 mm that corresponds to an angular distance of 0.07$^\circ$
on the sky.  It is small enough for the position determination
accuracy of 0.2$^\circ$ achieved.  The detector position resolution of
2.0 mm (FWHM) for 5.9-keV X-ray at the nominal anode voltage of 1650
was constant within the statistical uncertainty of 0.05 mm (1
$\sigma$) since the ground tests.

\subsection{Point Spread Function in the Anode-wire Direction}

The point spread function (PSF) of the GSC is determined by the
angular response of the slit-and-slat collimator and the position
response of the position-sensitive gas counter along the anode wire.
The collimator is designed to have an angular resolution of
1.0--1.5$^\circ$ in FWHM, which slightly depends on the X-ray incident
angle.  The position resolution of the gas counter depends on the
X-ray energy.  The resolution for X-rays with a normal incident angle
is inversely proportional to the number of multiplied signal charges
and varies from 4 to 1 mm in the GSC energy band from 2 to 30 keV
(\cite{mihara_paper1}).  For events with slant incident angles, the
variation of the photon-absorbed depth in a gas cell can also degrade
the position resolution.  The relation between the positional scale on
the detector and angular distance on the sky also depends on the
incident angle.  Thus, the PSF is represented by a complex function of
X-ray energy and incident angle.

\begin{figure}
  \begin{center}
    %\FigureFile(8.5cm,){psf_phidist_cphi5deg_v2.eps}
    %\FigureFile(8.5cm,){psf_phidist_cphi33deg_v2.eps}
    \FigureFile(8.5cm,){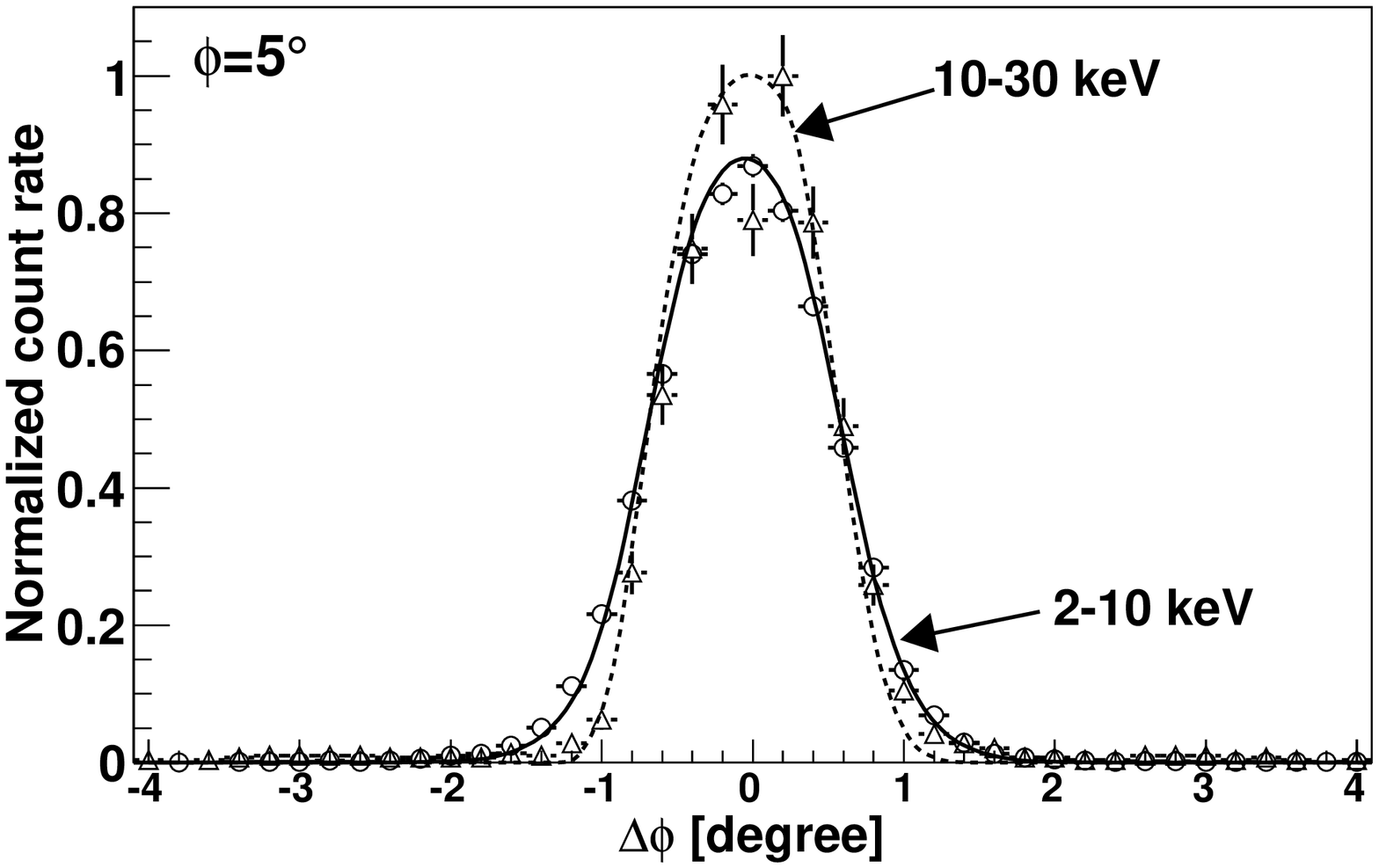}

    \FigureFile(8.5cm,){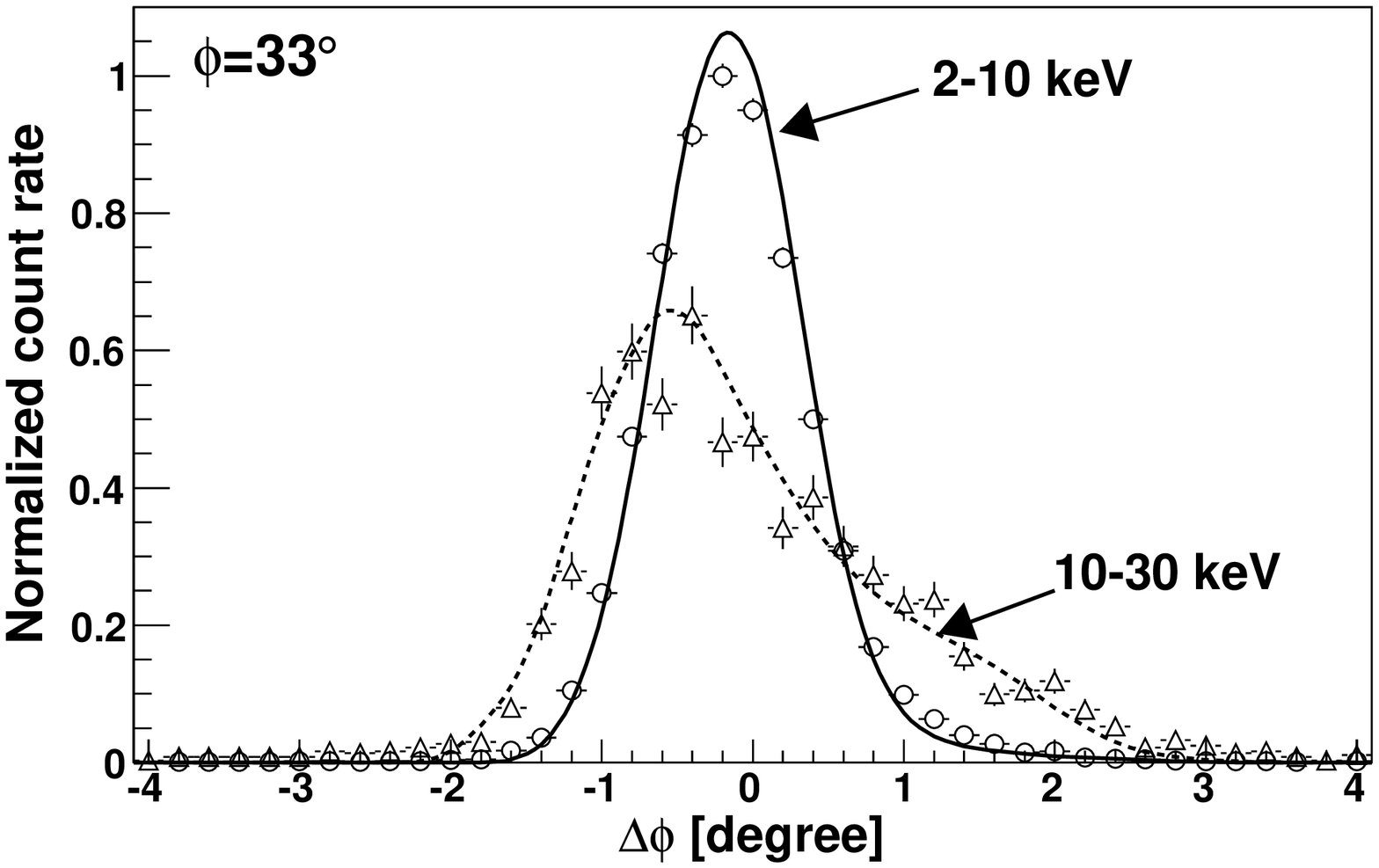}

  \end{center}
    \caption{ PSF cross-section profiles along the detector anode-wire
      direction at each source-incident angle, $\phi$, of $5^\circ$
      (top) and $33^\circ$ (bottom) in two energy bands of 2--10 keV
      (solid line) and 10--30 keV (dashed line).  The power-law energy
      spectrum of photon index, 2.1, is assumed.  Actual observed
      profiles of Sco X-1 taken at each source angle are plotted
      with 1-$\sigma$ error bars in the 2--10 keV (circle)
      and 10--30 keV (triangle) bands.  }
    \label{fig:gscpsfproj}
\end{figure}

Figure \ref{fig:gscpsfproj} illustrates the cross-section profiles of
PSF models for the direction along the detector anode-wire at each
photon-incident angle, $\phi$, of $5^\circ$ and $33^\circ$ in 2--10
keV (soft) and 10--30 keV (hard) energy bands (see the geometry in
Figure \ref{fig:maxi_gsc_schematic}).  We also show the actual
observed profiles taken from the data of a bright X-ray source, Sco
X-1, for comparison.  The PSF models agree with the data within the
statistical uncertainty.  For events of $\phi=5^\circ$, which is close
to the normal incident angle, the angular resolution in the hard band
is better than that in the soft band according to the better detector
position resolution along the anode wire for the higher-energy
X-rays. For events with a slant incident angle of $\phi=33^\circ$, the
variation of the absorbed depth yields the broad distribution along
the anode wire and degrades the resolution especially in the hard
band.

\subsection{Scan Transit Profile and PSF in the Scan Direction}
\label{sec:scan_transit}

\begin{figure}
  \begin{center}
    %\FigureFile(8.5cm,){trans_time_rot.ps} 
    \FigureFile(8.5cm,){fig8.ps} 

  \end{center}
    \caption{ Scan-transit time as a function of source-acquisition
      angle for each GSC unit of GSC\_0, GSC\_1, and GSC\_2, in the
      ISS normal attitude. }
    \label{fig:transitduration}

    %\vspace{3mm}

  \begin{center}
    %\FigureFile(8.5cm,){eahist_lc_5deg_v2.eps} 
    \FigureFile(8.5cm,){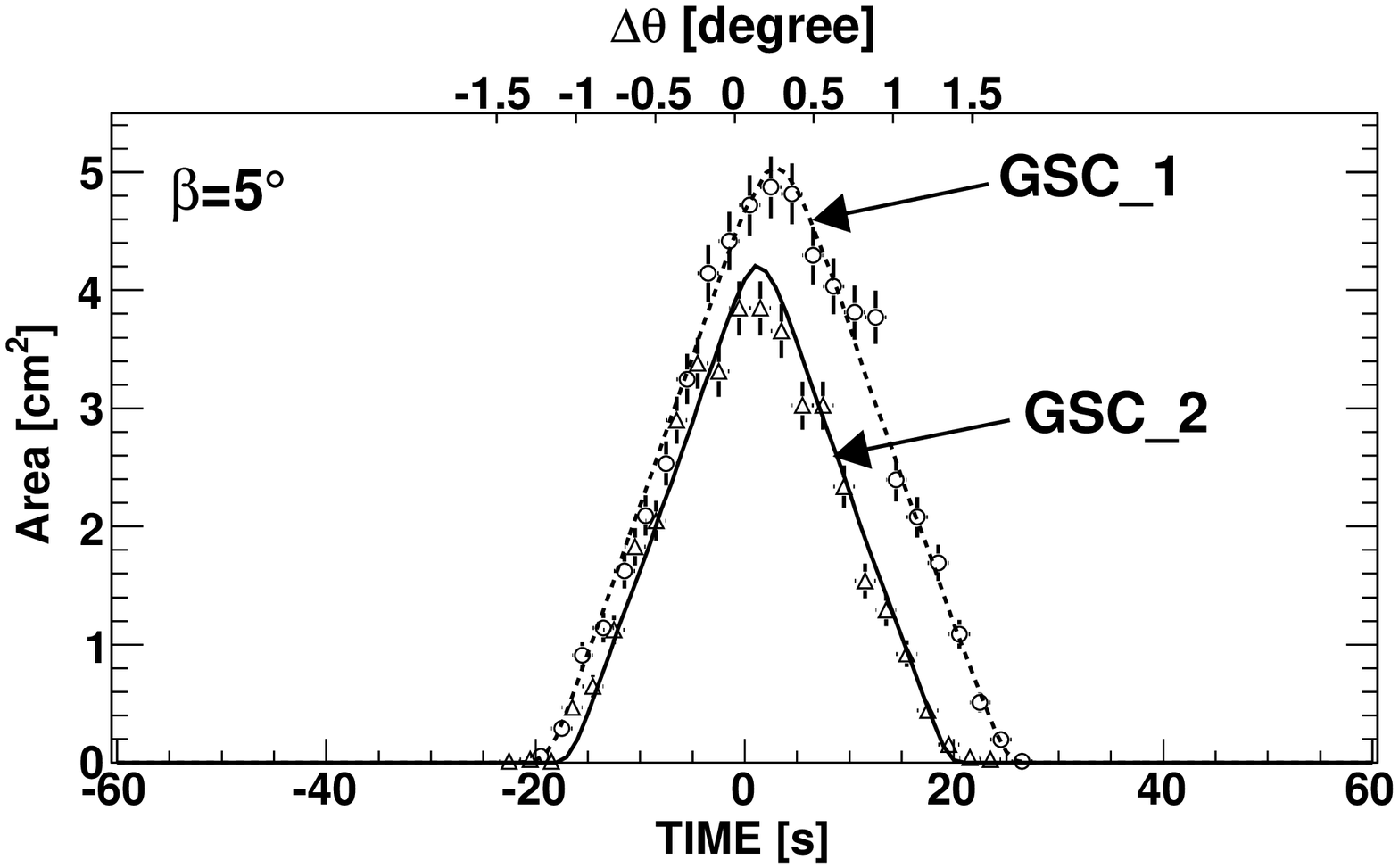} 

    %\vspace{2mm}

    %\FigureFile(8.5cm,){eahist_lc_72deg_v2.eps} 
    \FigureFile(8.5cm,){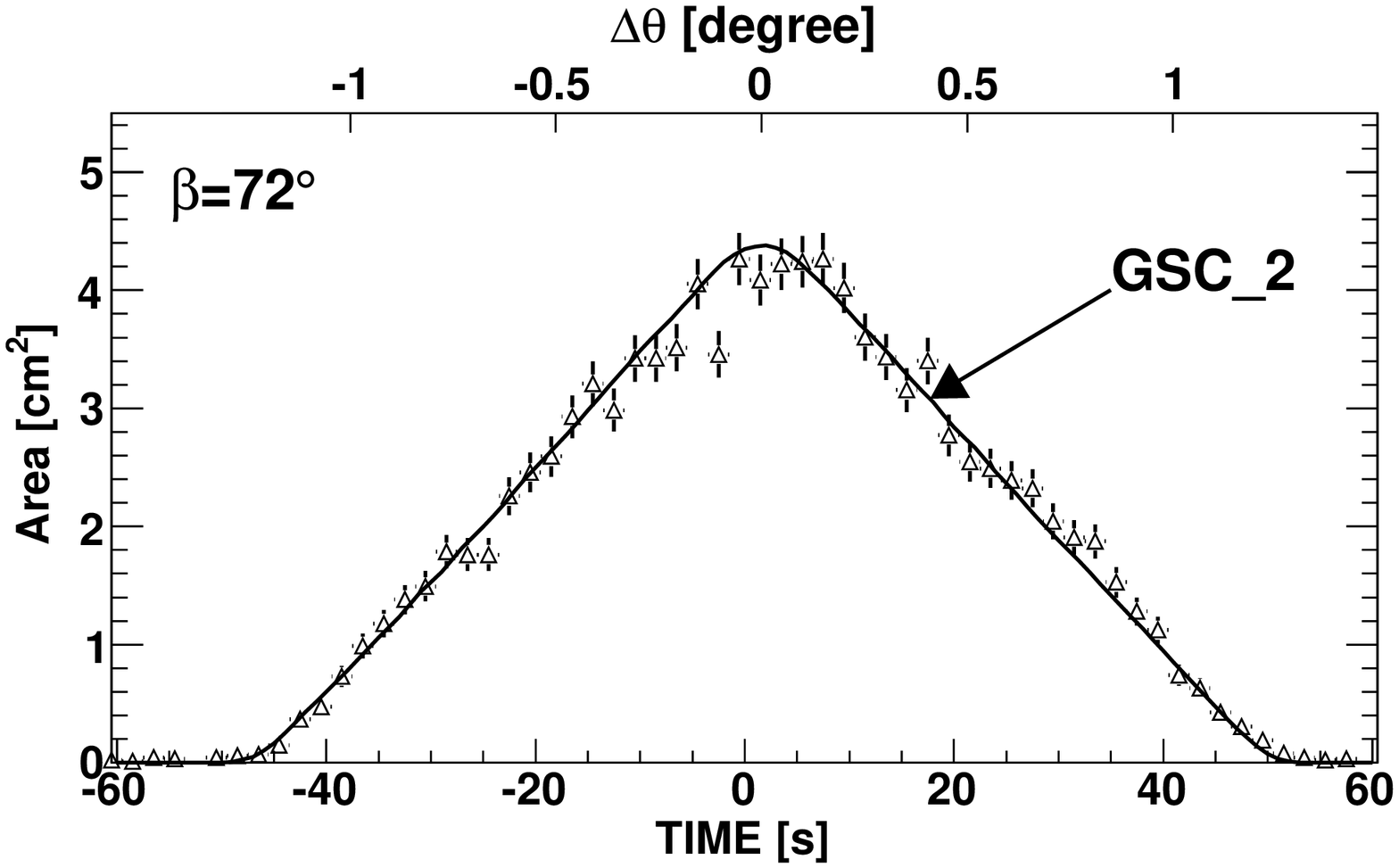} 

  \end{center}
    \caption{ Time variation of effective area during each scan
      transit at source-acquisition angle of 5$^\circ$ for GSC\_1 and
      GSC\_2 units (top) and 72$^\circ$ for GSC\_2 units (bottom). The
      angular scale of each transit is shown at the top of each panel.
      Observed raw event rates of Sco X-1 by each GSC unit are plotted
      together with 1-sigma error bars. }
    \label{fig:transitprofile}
\end{figure}

The GSC typically scans a point source on the sky during a transit of
40--150 seconds with the FOV of 1.5$^\circ$-width (FWHM) every
92-minute orbital period.  The transit time depends on the source
incident angle for each GSC counter $\phi$ as well as the acquisition
angle from the ISS rotation equator $\beta$ (see the geometry in
figure \ref{fig:maxi_gsc_schematic}).  Figure
\ref{fig:transitduration} illustrates the relation between the
scan-transit time on each GSC counter and the source acquisition angle
from the rotation equator in the ISS normal attitude. The scan
duration increases from the ISS rotation equator ($\beta=0^\circ$)
towards the pole ($\beta=\pm 90^\circ$) in the side cameras such as
GSC\_0 and GSC\_2 according to the FOV-area angle from the rotation
pole.

The detector area for the target changes according to the triangular
transmission function of the collimator during each transit.  It
modulates time variations of observed photon count rates and
represents the PSF in the scan direction if the source flux is
constant.  The scan direction is perpendicular to the detector
anode-wire direction in the normal ISS attitude.  Figure
\ref{fig:transitprofile} shows time variations of effective areas
during scan transits at typical two source-acquisition angles of
5$^\circ$ and 72$^\circ$, where the angular scales of the collimator
transmission function are also indicated at the top on each panel.
The observed count rates of Sco X-1 taken at the same acquisition
angle are plotted together for comparison.  The calculated scan
duration and the observed period are agreed well with each other.

\subsection{Exposure Map and Sensitivity}

\begin{figure}
  \begin{center}
    %\FigureFile(8.5cm,){expmap_1orb_equ2.eps}
    %\FigureFile(8.5cm,){expmap_1day_equ2.eps}
    %\FigureFile(8.5cm,){expmap_27day_equ2.eps}
    \FigureFile(8.5cm,){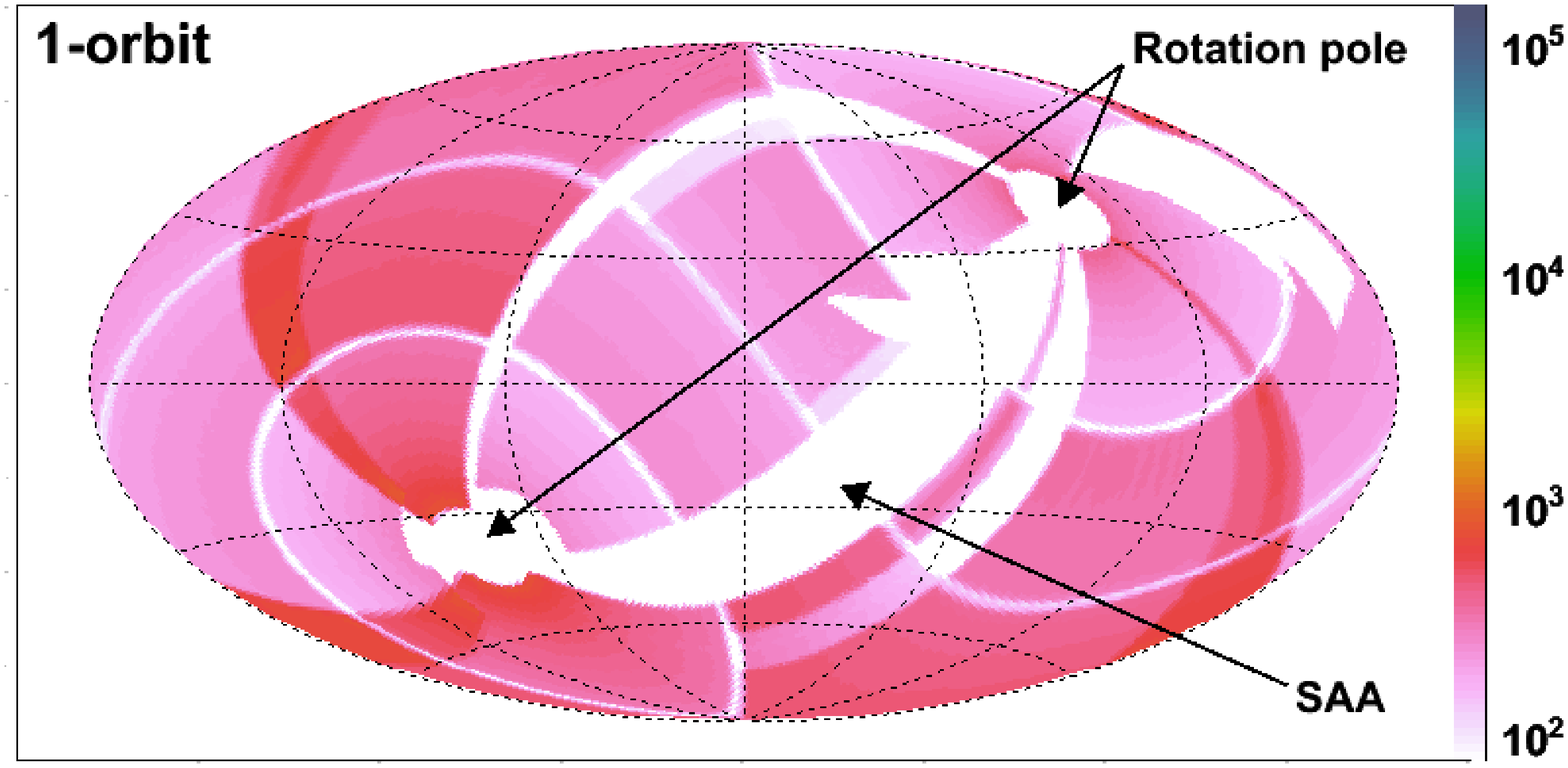}
    \FigureFile(8.5cm,){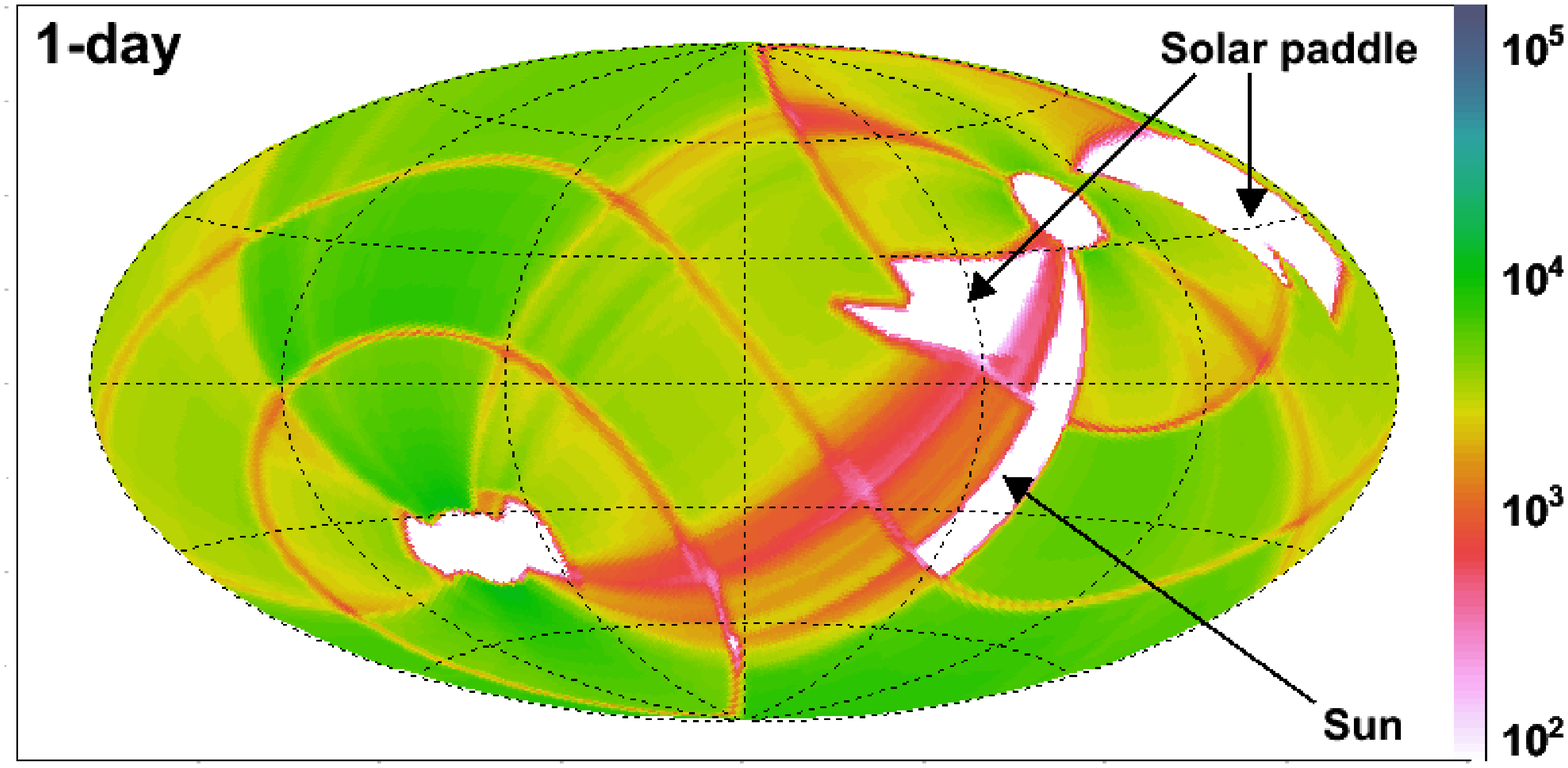}
    \FigureFile(8.5cm,){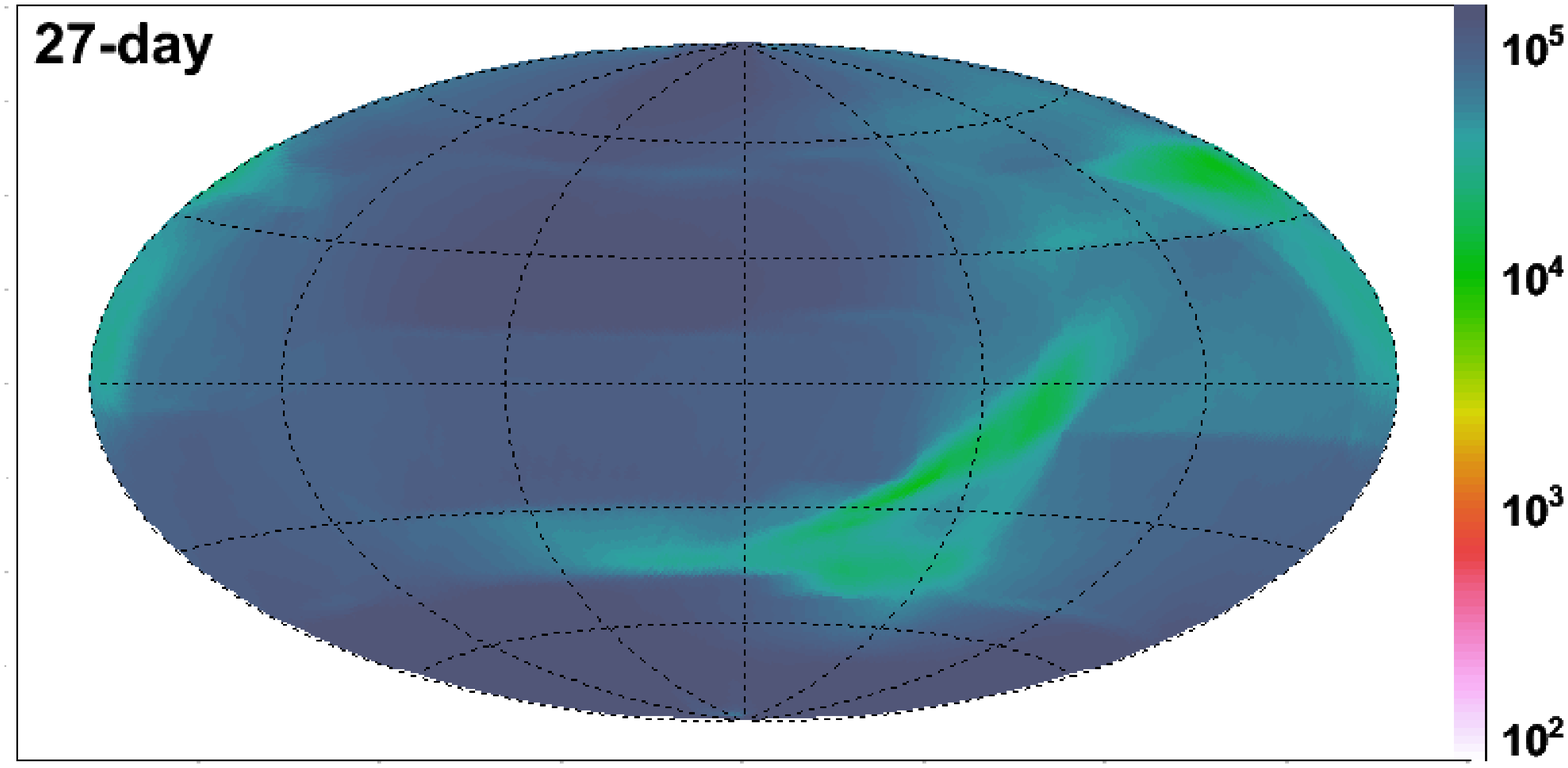}
  \end{center}
    \caption{ GSC exposure map by scans of 92-minute orbital period
      (top), 1 day (middle), and 27 days (bottom) since 2010 January 2
      00:00 (UT).  The color scales of all three panels are identical
      and represent the effective exposure for the sky direction in
      the units [cm$^2$s].  The uncovered areas for the rotation pole,
      the Sun direction, the SAA, and the interference with the solar
      paddles are indicated with arrows.  }
    \label{fig:allskyea}

\end{figure}

An exposure map on the sky for a given period is calculated from the
counter HV-operation history, the detector effective area as a
function of a photon incident angle, and information of the ISS orbit,
attitude, and configuration parameters interfering the FOV such as the
solar paddles, the space-shuttle vehicle docked on the ISS.  Figure
\ref{fig:allskyea} shows the actual exposure maps by all operated GSC
counters for 92 minutes of an ISS-orbital cycle, one day, and 27 days
since January 2, 2010, 00:00 (UT), calculated from all the required
information during these periods.  GSC covers approximately 85\% of
the whole sky for one orbit except for the orbits including the SAA
passage, and 95\% for a day.  The daily map has uncovered area for the
Sun direction, the solar-paddle shadow, and the rotation pole that
drift on the sky according to the precession of the ISS orbit.  We can
achieve the 100\% full coverage in every three weeks.

An actual exposure time for an arbitrary direction on the sky was
typically 4000 cm$^2$s per day.  It is about one third of the initial
expectation because the HV operation at the high latitude above
40$^\circ$ was stopped and four counters out of 12 were disabled in
these period.  The daily 5-$\sigma$ source sensitivity is expected to
become 15 mCrab, which is three times worse than that of the
pre-flight simulation (\cite{matsuoka_pasj2009}).

\section{Calibration Status}

We here present the status of the GSC response calibration for data
taken at the nominal HV of 1650 V.  We also exclude data detected by
anodes \#1 and \#2 of all GSC counters because some of them show
complicated energy-PHA relations at low PHA, which is probably due to
feedback crosstalks on the circuit board.  The responses functions for
the data taken at the reduced HV of 1550 V and by the anode \#1 and
\#2 are being developed and will be updated soon.

\subsection{Alignment and Position Localization Accuracy}

The attitude of the MAXI payload module is continuously measured with
the on-board Attitude Determination System (ADS), which consists of a
Visual Star Camera (VSC) and a Ring Laser Gyroscope (RLG). Its
accuracy is estimated to be better than an arc minute from the ground
calibration tests (\cite{horike_2009}).  The source location of an
incident X-ray on the sky is calculated from the position coordinates
on the detector where the X-ray is absorbed and the attitude
parameters at the event time.  The alignments of the collimators and
detectors on the payload module are calibrated using standard X-ray
sources whose positions and intensities are well-known.  The accuracy
of 0.2$^\circ$ in the 90\% containment radius of the best determined
position has been confirmed so far using several bright-source samples
selected randomly \citep{morii_phye2010}.

\subsection{Effective Area and Light curve}
\label{sec:eff_lc}

The visibility and the area of each GSC unit for a given X-ray source
on the sky are always changing according to the ISS orbital motion and
interferences with other relevant ISS activities, as shown in section
\ref{sec:scan_transit}.  To derive the intrinsic time variation of a
certain X-ray source, the time-dependent effective area for the target
has to be known.  The light-curve response builder calculates the area
curve of each GSC unit from the source coordinates, the ISS location,
attitude, and the area of the GSC unit as a function of the source
incident angle.  The area of all the GSC units were calibrated on the
ground using pencil X-ray beams \citep{morii_spie2006}.

\begin{figure}
  \begin{center}
    %\FigureFile(8.5cm,){crab_lc_2bin_4-10kev_v2.eps}
    \FigureFile(8.5cm,){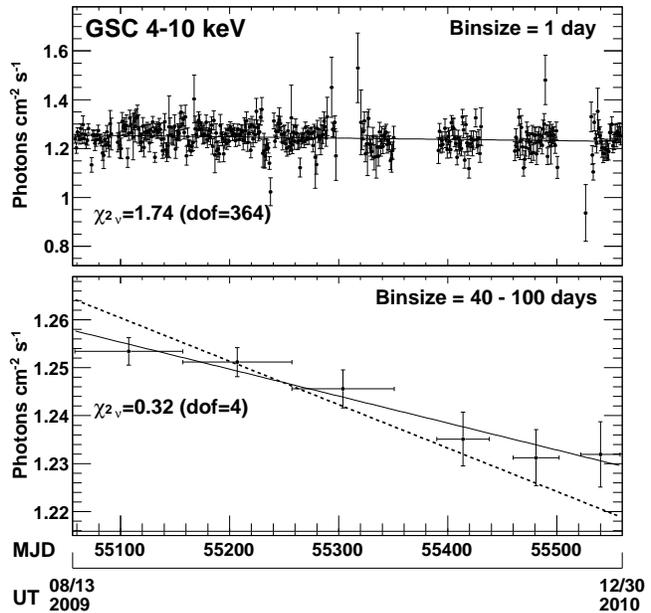}
  \end{center}
    \caption{ (Top) Crab nebula light curve by GSC in 4--10 keV band
      in one-day time bin.  The vertical errors represent the 1-sigma
      statistical uncertainties.  Solid line represents the best-fit
      linear model.  The reduced chi-square ($\chi^2_{\nu}$) and the
      degrees of freedom (DOF) of the fit is shown on the panel.
      (Bottom) The same Crab nebula light curve but in 40--100-day time
      bin and the best-fit linear model.  The model with a slope
      extrapolated from the 5.1\% decline from MJD 54690 to 55435
      observed by RXTE/PCA in 2--15 keV band
      \citep{2010arXiv1010.2679W} is shown together with dashed line.}
     \label{fig:crab_lightcurve} 
\end{figure}

The top panel of figure \ref{fig:crab_lightcurve} shows the obtained
light curve of Crab nebula in the 4--10 keV energy band since the
mission start on August 15, 2009.  The data are collected from the
area-corrected photon count rates for each day.  Backgrounds are
estimated from the count rates for the adjacent sky region, which
corresponds to the data obtained from the successive time periods.  A
power-law energy spectrum with a photon index, $\Gamma=2.1$, is
assumed in the effective-area correction.  The light curve should be
approximately constant and consistent with the flux of the standard
Crab model.  The measured average flux over the entire period, $1.25$
photons cm$^{-2}$ s$^{-1}$, agrees with the standard Crab value
(\cite{2005SPIE.5898...22K}; \cite{2010ApJ...713..912W}).  However,
the fit to a linear model results in a reduced chi-squared
$\chi^2_\nu=1.74$ for 364 degrees of freedom (DOF), which is not
accepted within the 90\% confidence limit.  It is due to the limit on
the effective-area calibration accuracy on the daily basis.  The error
on the daily flux in the 4--10 keV band is estimated to be $3$\% with
the 1-$\sigma$ range.  The calibration over the entire 2--30 keV
energy range is in progress.

Recently, \citet{2010arXiv1010.2679W} presented evidences of flux
variations of Crab nebula by $\sim$7\% on a 3-year timescale,
consistently observed by Fermi/GBM, RXTE/PCA, Swift/BAT, and
INTEGRAL/ISGRI.  The flux in 2--15 keV band observed by RXTE/PCA
exhibits a 5.1\% decline from MJD 54690 to 55435.  We examined the
variation with the GSC data.  The bottom panel of figure
\ref{fig:crab_lightcurve} shows the variation of the obtained GSC
4--10 keV flux by each interval of about 100 days.  The start and the
end of each interval are slightly adjusted according to the
observation-time coverage.  A continuous decline is seen clearly.  The
fluxes averaged by the long term are derived from large event data
taken by various camera angles and detector positions.  Therefore, the
calibration uncertainty of the effective area on daily basis, 3\% in 1
$\sigma$, is supposed to be smeared out in it and the observed decline
is considered to be real.

We fitted the data to a linear function and obtained the decline ratio
of $2.2\pm 0.5$\% (1-$\sigma$ statistical uncertainty) from MJD 55058
to 55558.  The ratio is slightly smaller than that extrapolated from
the 5.1\% decline in MJD 54690--55435 observed by the RXTE/PCA in
2--15 keV band.  We tried to fit the GSC data to a linear model with
the fixed slope determined by the RXTE/PCA, as shown in figure
\ref{fig:crab_lightcurve}.  The fit is marginally unacceptable with
$\chi^2_{\nu}=1.92$ for 5 DOF.  It might imply the dependence of the
variation amplitude on the energy band, as reported in
\citet{2010arXiv1010.2679W}.

\subsection{Energy Response Matrix}

The energy response matrix is represented by a product of the
effective area and the dispersion relation between the incident X-ray
energy and the PHA of the output signal.  The dispersion relation as
well as the gas gain depend on the 3-dimensional location where the
X-ray photon is absorbed in a detector gas cell.  The spatial gain
variation and the time variation in orbit are corrected event by event
in the first data-reduction process using calibration data collected
from the ground and in-orbit test results.  All the dispersion
relations required to derive the response matrix were taken in the
ground test (\cite{mihara_spie2002}).

The energy response builder calculates the effective area and the
energy redistribution matrix for a target source at given sky
coordinates during a given observation period. The program accumulates
instantaneous response functions for the source location according to
the variation of the incident angle during the active observation
period.

\begin{figure}
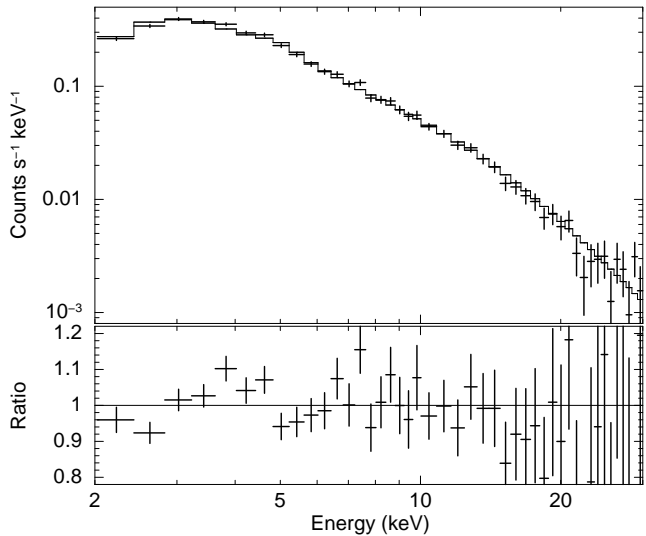

  \begin{center}
    %\FigureFile(8.5cm,){crab_spec_cor8_nhfix_rot.eps} 
    \FigureFile(8.5cm,){fig12.eps} 
  \end{center}
    \caption{ GSC Crab-nebula spectrum taken for a day on August 15,
      2009 (UT) and the best-fit power-law model with a fixed
      absorption of $0.35\times 10^{22}$ cm$^2$ (top). Data-to-model
      ratio (bottom).  }\label{fig:crab_spec}
\end{figure}

Figure \ref{fig:crab_spec} shows a Crab spectrum by GSC taken for a
day on August 15, 2009 (UT) and the best-fit power-law model with an
interstellar absorption, folded by the response matrix obtained by the
response builder.  Backgrounds are extracted from events in the
adjacent sky region.  The best-fit parameters are summarized in table
\ref{tab:crab_spec_fit}.  The model-to-data ratio shows that the
obtained spectrum is well reproduced within 10\% over the 2-30 keV
energy range.  The best-fit absorption column density, $0.72\pm0.22
\times 10^{22}$ cm$^{-2}$ is slightly higher than the standard Crab
value, $0.35 \times 10^{22}$ cm$^{-2}$ (\cite{2005SPIE.5898...22K})
although the fit with the absorption fixed at the standard value is
also accepted within the 90\% confidence limit.  The discrepancy is
considered to be in inaccuracy of the effective-area calibration at
the low energy band.
The fine calibration is in progress.

\begin{table*}
\caption{Best-fit parameters for GSC Crab-nebula spectrum}
\begin{center}
\label{tab:crab_spec_fit}
\begin{tabular}{lcccc}
\hline
\hline
Model  & $N_{\rm H}$$^{\rm a}$  & $\Gamma$$^{\rm b}$ & Norm.$^{\rm c}$ & $\chi^2_\nu$ (DOF) \\
\hline 
Power-law, $N_{\rm H}$: free  ~~~ &  $0.72\pm 0.22$ & $2.15\pm 0.05$ & $11.4\pm 1.0$ & 1.08 (43) \\
Power-law, $N_{\rm H}$: fixed ~~~ &  0.35 & $2.08\pm 0.03$ & $9.9\pm 0.4$ & 1.25 (44) \\
\hline
\end{tabular}

\begin{tabular}{l}
All errors represent 90\% confidence limits of statistical uncertainty.\\
$^{\rm a}$ Absorption hydrogen column density in units of $10^{22}$ cm$^{-2}$. \\
$^{\rm b}$ Power-law photon index \\
$^{\rm c}$ Power-law normalization in units of photons cm$^{-2}$ s$^{-1}$ keV$^{-1}$ at 1 keV.\\
\end{tabular}
\end{center}
\end{table*}

\subsection{Timing}

\begin{figure}
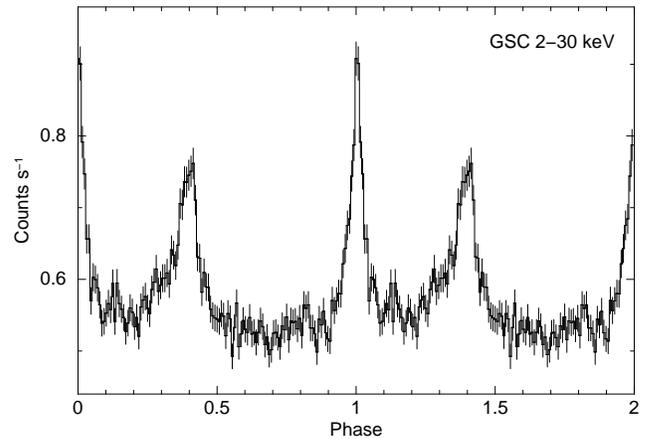

  \begin{center}
    %\FigureFile(8.5cm,){efold_mjd55197_55202_norm0_fig_rot.ps} 
    \FigureFile(8.5cm,){fig13.ps} 
  \end{center}
    \caption{ Folded light curve of the Crab pulsar in 2--30 keV band
      obtained with GSC during a period from January 1 to January 5,
      2010.  The vertical count rate represents the event rate from
      the target direction over the entire elapsed period.  }
    \label{fig:crab_pulse_profile}

\end{figure}

In-orbit timing calibrations are performed using signals from X-ray
pulsars or binaries, such as Crab pulsar and Cen X-3 with 33
millisecond and 4.8 second period, respectively.  Events within the
PSF of these pulsars are extracted and applied to the barycentric
correction. The pulsation of Crab pulsar can be detected for every
one-day interval.  Figure \ref{fig:crab_pulse_profile} show the folded
pulse profiles since January 1 to January 5, 2010.  The absolute
timing and the relative stability are estimated by the relative pulse
phase of the Crab pulsar to those with the radio wavelength
\citep{1993MNRAS.265.1003L} and RXTE/PCA \citep{2004ApJ...605L.129R},
and comparison of the Cen X-3 pulse period and phase with those of the
Fermi/GBM \citep{meegan2009}.  We confirmed the timing stability of an
order of $10^{-9}$.

\section{All-sky image}

\begin{figure*}
  \begin{center}
    %\FigureFile(15cm,){maxi_gsc_allskyimg_wb.eps}
    \FigureFile(15cm,){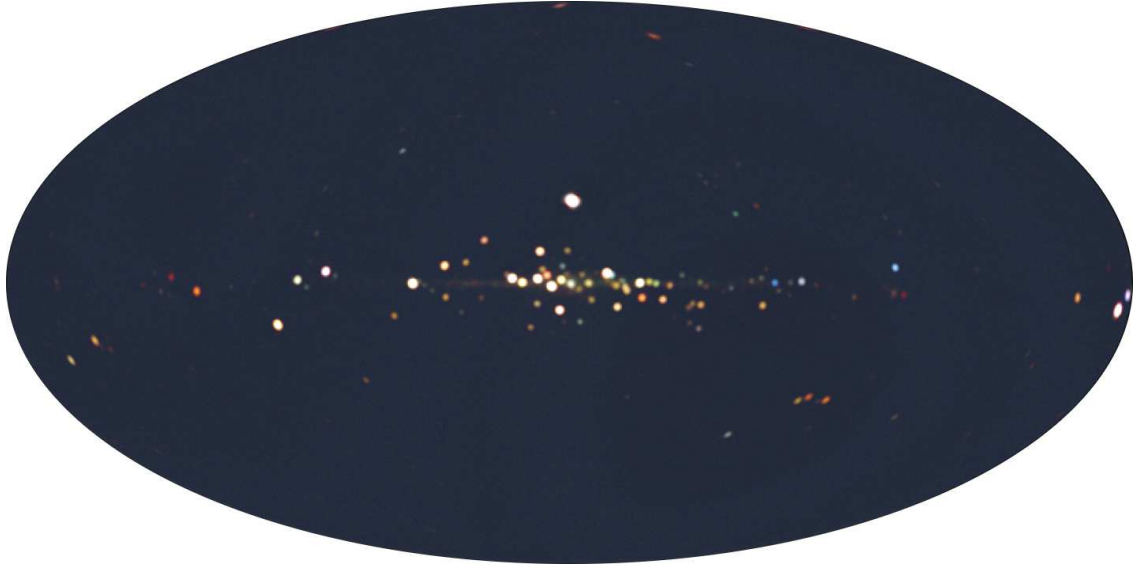}
  \end{center}
    \caption{ GSC all-sky image taken for the first year.  Red, green,
      and blue color maps are scaled logarithmically and represent the
      intensities in 2-4, 4-8, and 8-20 keV bands, respectively.  }
    \label{fig:gsc_allskyimg}
\end{figure*}

Figure \ref{fig:gsc_allskyimg} shows the GSC all-sky X-ray image taken
for the first year.  The red, green, and blue color scales represent
the intensity in 2-4, 4-8, and 8-20 keV energy band, respectively.
The diameter of point source images roughly represents the relative
brightness according to the PSF spread.  We can easily see more than a
hundred of discrete X-ray sources over the whole sky as well as
unresolved Galactic ridge emission along the Galactic plane with a
scale height of 1--2$^\circ$.

\section{Summary}

The GSC in-orbit performance on the MAXI mission payload is verified
from the initial commissioning operation. Table
\ref{tab:performance_summary} briefly reviews the relevant parameters.
They are summarized by the following points.

\begin{enumerate}

\item The optimal configurations of the read-out electronics and
  on-board data processing at the nominal high voltage of 1650 V are
  established. The gas gain and the amplifier gain are confirmed to
  agree with those expected from the ground calibration tests.  The
  gain change during the first year is confirmed to be less than 1
  \%, which is small enough for the requirements of the calibration
  accuracy.

\item Two gas counters out of the twelve were turned off for the
  high-voltage breakdown during the first two-months commissioning
  operation and another breakdown occurred in the ninth month.  These
  breakdown would be presumably caused by repeated discharges.  The HV
  operation under the high particle-background, which includes area at
  the high latitude above 40$^\circ$ as well as the SAA, were stopped
  since September 27, 2009 in order to avoid any damage that can
  develop into the discharges.  The three counters suffering from the
  breakdown are now operated with a limited sensitivity using
  alive-anode partitions.

\item The background cosmic-ray rate is consistent with that
  extrapolated from the past experiments in the low-earth orbit using
  the relation with the geomagnetic COR.  The efficiency of the
  on-board background rejection by the anti-coincidence-hit logic is
  also found to be comparable to those of the similar instruments,
  Ginga-LAC.

\item The PSF with an FWHM of 1.5$^\circ$, the scan transit time from
  40 to 150 seconds per orbital cycle, the all-sky coverages of 85\%
  per 92-minute orbital period and 95 \% per day are verified, which
  agrees with the pre-flight design.  The effective exposure time for
  a celestial target is typically 4000 cm$^{2}$ s.  The daily
  5-$\sigma$ source sensitivity is expected to be 15 mCrab in the
  condition.
 
\item The instrument response in orbit has been calibrated using Crab
  nebula data.  The fitting results verified that the uncertainty of
  the effective area and the energy response would be less than 10\%
  if data of a few anodes suffering from cross talk on the
  circuit board are ignored.  The relative timing accuracy was also
  confirmed to be normal.

\end{enumerate} 

\begin{table*}
\caption{Summary of GSC in-orbit performance}
\begin{center}
\label{tab:performance_summary}
\begin{tabular}{ll}
\hline
%Detector position resolution & 2 mm at 5 keV\\
Spatial resolution            & 1.5$^\circ$ (FWHM)\\
Source localization accuracy  & 0.2$^\circ$ (90\% containment radius)\\
All-sky coverage              & 85\% per orbit (nominal 92 minutes), 95\% per day \\
Exposure                      & 4000 cm$^{2}$ s per day (typical)\\
Energy resolution             & 18\% (1$\sigma$) at 5.9 keV\\
Effective area accuracy       & 3\% (1$\sigma$) at 4--10 keV\\
%Time accuracy                 & xx\\ 
Sensitivity                   & 15 mCrab per day in 2--30 keV (typical)\\
\hline
\end{tabular}

\begin{tabular}{l}
%$^{\rm a}$ The ISS orbital period is 92 minutes. \\
\end{tabular}
\end{center}
\end{table*}

%\section*{Acknowledgments}
\bigskip

We are grateful to Mark Finger and Fermi/GBM pulsar project team for
providing the pulsar ephemeris data for timing calibration.  We also
thank Keith Jahoda for stimulating and constructive comments. This
research was partially supported by the Ministry of Education,
Culture, Sports, Science and Technology (MEXT), Grant-in-Aid for
Science Research 19047001, 20244015, 21340043, 21740140, 22740120 and
Global- COE from MEXT ``Nanoscience and Quantum Physics'' and ``The
Next Generation of Physics, Spun from Universality and Emergence''.

%%%
% See the manual for the detail.
%%%

\end{document}